\RequirePackage{fix-cm}
\documentclass[smallextended]{svjour3}       
\smartqed  
\usepackage{graphicx}
\usepackage{caption}
\usepackage{multirow}
\usepackage{array}

\usepackage{amsmath}
\usepackage{algorithm}
\usepackage{algorithmicx}
\usepackage{algpseudocode}
\usepackage[numbers]{natbib}

\newcommand{\tabincell}[2]{\begin{tabular}{@{}#1@{}}#2\end{tabular}}
\begin{document}

\title{SPLZ: An Efficient Algorithm for Single Source Shortest Path Problem Using Compression Method
}


\author{Jingwei Sun         \and
        Guangzhong Sun 
}


\institute{Jingwei Sun \at
              School of Computer Science and Technology, University of Science and Technology of China \\
              \email{sunjw@mail.ustc.edu.cn}           
           \and
           Guangzhong Sun \at
              School of Computer Science and Technology, University of Science and Technology of China\\
              \email{gzsun@ustc.edu.cn}
}

\date{Received: date / Accepted: date}

\maketitle

\begin{abstract}
Efficient solution of the single source shortest path (SSSP) problem on road networks is an important requirement for numerous real-world applications. This paper introduces an algorithm for the SSSP problem using compression method. Owning to precomputing and storing all-pairs shortest path (APSP), the process of solving SSSP problem is a simple lookup of a little data from precomputed APSP and decompression. APSP without compression needs at least 1TB memory for a road network with one million vertices. Our algorithm can compress such an APSP into several GB, and ensure a good performance of decompression. In our experiment on a dataset about Northwest USA (with 1.2 millions vertices), our method can achieve about three orders of magnitude faster than Dijkstra algorithm based on binary heap.
\end{abstract}
\newpage
\section{Introduction}
\label{sec:introduction}
The single source shortest path (SSSP) problem is a classic algorithm problem, and is also a model for numerous real-world applications, such as navigation, facilities location, logistics planning. Generally, given a graph $G = (V, E)$, and a source vertex $s$, the goal of SSSP problem is to find the shortest paths from $s$ to all other vertices in the graph.

Effective precomputation plays an important role of many efficient algorithms of SSSP problem. This kind of algorithms contains two phases: precomputing some supporting information offline, and computing the final results online. A straightforward precomputation method is precomputing all-pair shortest path (APSP) and storing it in memory. Then the time complexity of SSSP problem online is $O(\|V\|)$ for a simple lookup and output, However, space consumption for raw data of APSP needs at least $O(\|V\|^2)$. There is no sufficient memory to run this algorithm on an ordinary machine for a large-scale graph. For example, a graph with one million vertices needs at least 1 TB memory. It is an extreme situation that we only record the shortest path tree. An APSP includes $O(\|V\|)$ trees, and each tree needs $O(\|V\|)$ bytes to record the parent of every vertex. In road networks, the degree of a vertex is lower than 255, so recording the parent of a vertex needs 1 byte. At this situation, the APSP of one million vertices takes about 1TB.

In this paper, we propose a compression method to reduce the space cost of precomputation effectively, and ensure a linear time complexity for decompressing. We call our method shortest path with Lempel Ziv (SPLZ), which means it is a modification of LZ77\cite{ziv1977universal} algorithm. However, the offline time complexity is $O(\|V\|^2)$, so SPLZ is not very suitable to work on a continental-sized graph, for example, the Europe road network with 18 million vertices. The number of vertices of our experiment road networks is from 300,000 to 1,200,000. Though they are not continental-sized, they may still be representations of large-scale road network in real-world applications. SPLZ could work well at this scale.


In one of our experiment on road network of Northwest USA (about 1.2 million vertices), SPLZ can compress the APSP of this graph (about 1.4 TB) into several GB. It is affordable for a high-end PC or an ordinary workstation. If memory is insufficient, we can store the compressed APSP into extended memory. Our experiments show that using extended memory can still achieve a good performance of decompressing.

There are three main contributions of our paper:
\begin{itemize}
\item  We design an effective compression scheme for storing APSP data. With this method, we can take full advantages of information generated by precomputation. When memory is not enough, we can store the compressed APSP into extended memory and still keep a good performance of decompression.
\item  We develop a fast algorithm named SPLZ to solve SSSP problem. Our algorithm on single core can achieve about three orders of magnitude faster than Dijkstra algorithm based on binary heap. This performance is about two to three times as much as the time cost of copying an array of length $\|V\|$ using standard C library function memcpy().
\item  SPLZ is simple to be implemented. SPLZ does not use complex data structures and elaborate skills. In the offline phase, SPLZ uses Dijkstra (or PHAST) and LZ77 with a little modification. In the online phase, the operation of SPLZ for solving SSSP is copying an array of length $O(\|V\|)$.
\end{itemize}
The remainder of this paper is organized as follows: Section \ref{sec:Related works} describes related works. Section \ref{sec:Basic SPLZ} introduces the basic idea of SPLZ. Section \ref{sec:Improvements} details the implement of SPLZ. Section \ref{sec:Experiment} reports the experimental results. Conclusion is made in Section \ref{sec:Conclusion}.
\section{Related works}
\label{sec:Related works}
SSSP problem has been widely researched. Dijkstra\cite{dijkstra1959note} algorithm is the most classic method for SSSP problem. To improve the performance of Dijkstra, researchers have adopted numerous type of priority queue.  It has a series of modification like DIKB\cite{dial1969algorithm}, DIKBD\cite{cherkassky1996shortest}, DIKH\cite{cormen2001introduction}, DIKR\cite{ahuja1990faster}. They are also called label setting algorithm. Another classic algorithm for SSSP problem is Bellman-Ford\cite{bellman1956routing}. It is classified to another type of algorithm called label correcting algorithm. Besides Bellman-Ford, label correcting algorithm includes many others like PAPE\cite{pape1974implementation}, TWO-Q\cite{pallottino1984shortest}, THRESH\cite{glover1985new}, SLF\cite{bertsekas1993simple}. They are based on different label correcting strategies. Both label setting algorithm and label correcting algorithm can be described by a unified framework\cite{gallo1986shortest}. These algorithms are designed for general purpose, so they do not have preprocess phase and other optimization skills for road network. These algorithms usually perform not so well for SSSP problem on large scale road network.

Some algorithms accelerate solving the SSSP problem with parallelism. The classic parallel algorithms for SSSP problem contain parallel asynchronous label correcting method\cite{bertsekas1996parallel} and $\Delta$-stepping\cite{meyer2003delta}. Goldberg et al\cite{cherkassky2009shortest} pointed out that traditional parallel algorithm for SSSP problem usually do not take full advantages of modern CPU architecture, like multi-core, SSE. $\Delta$-stepping has less acceleration on large-scale road network\cite{madduri2006parallel}. In 2011, Delling and Goldberg et al developed PHAST\cite{delling2013phast}, which is the fastest algorithm at present. PHAST makes full use of SSE, multi-core, and is elaborately designed to obtain a low cache missing. Its performance of GPU modification on large scale road network is up to three orders of magnitude faster than Dijkstra on a high-end CPU. Using PHAST instead of Dijkstra can reduce the time cost of offline phast of SPLZ. Parallel technique is also used in SPLZ to accelerate the offline precomputing phase.

Precomputing methods replace the time consumption online with time and space consumption offline. It is a efficient approach for solving shortest path problem on large-scale graph. Many high performance algorithm for point-to-point shortest path problem have a precomputing process, including Highway Hierarchies\cite{sanders2005highway}, Transit Node Routing\cite{bast2007fast}, ALT\cite{goldberg2005computing} et al. In 2008, Geisberger proposed Contraction Hierarchies\cite{geisberger2008contraction} algorithm. It is not only a good algorithm for calculating the answer of shortest path, but also a efficient method for precomputing. Many outstanding algorithms like Transit Node Routing\cite{arz2013transit}, PHAST and Hub-based labeling\cite{abraham2011hub, Abraham:2012:HLS:2424321.2424365} adopted it as precomputing method. Hub-based labeling is the state-of-the-art point-to-point shortest path algorithm. It takes more space than others to obtain the fastest online performance. Similarly, SPLZ precomputes much more information (the total APSP) than these algorithms, using more offline time and space consumption. Thus SPLZ can achieve high online performance.

Some algorithms have good performance online, but space consumption of precomputing is too huge to be loaded in memory. Compression methods are practical way to reduce the space consumption. SILC\cite{sankaranarayanan2005efficient}, PCPD\cite{sankaranarayanan2009path}, and CPD\cite{botea2013moving} use compression method to reduce space complexity of storing APSP. These methods are to solve point-to-point shortest path problem on spatial networks (graphs where each node is labeled with coordinates). They store APSP in the form of ¡°first move table¡±. First move table can obtain high compression ratio by taking the advantage of path coherence\cite{sankaranarayanan2005efficient}. Graph partitioning usually is used for reducing online space in some algorithms, like Arc-flag\cite{hilger2009fast}, PCD\cite{delling2013customizable} and CRP\cite{Maue:2010:GSQ:1498698.1564502}. But for SILC et al, graph partitioning is used for clustering similar data in APSP. SPLZ has a similar preprocess like SILC et al, but SPLZ is used for SSSP problem, not point-to-point, so SPLZ adopts different compression and decompression method. Detailed difference will be described in section 3.1.

\section{Basic SPLZ}
\label{sec:Basic SPLZ}
\subsection{Main idea}
The main idea of SPLZ is to precompute APSP and then compress it for online lookup. In SPLZ, APSP is stored in the form of shortest path tree(SPT). Here SPT is an array of length $\|V\|$, which records the last move of the shortest path from a source vertex to all other vertices. For example, if we want to calculate the shortest path from vertex $v$ to other vertices, we will present the result by $SPT(v)$, which is a tree with root $v$. If $SPT(v)[u]=w$, it means that the predecessor of vertex $u$ along the shortest path from $v$ to $u$ is vertex $w$. Edge $(w,u)$ is the last move from $v$ to $u$, so SPT is also called last move table. Traditional algorithm for SSSP problem like Dijkstra usually present the result in the form of SPT too. If we store APSP straightforwardly, the space complexity is $O(\|V\|^2)$, for we need to store $\|V\|$ SPTs.

Path coherence described in \cite{sankaranarayanan2005efficient} reveals that ``vertices contained in a coherent region share the first segment of their shortest path from a fixed vertex''. For last move, path coherence still makes sense with a bit change on description: vertices contained in a coherent region share the last segment of their shortest path to a fixed vertex. Strictly, path coherence only holds when the fixed vertex is sufficiently far away. In large-scale road network, there always are numerous vertices which are sufficiently far away from each other, so path coherence holds in most cases. A related experiment is introduced in section 4.1 to show how frequently it holds. Path coherence implies that the data among multiple SPTs contain a large number of reduplicative sequences. This feature will lead to a high compression ratio with LZ-family algorithm.

SILC\cite{sankaranarayanan2005efficient}, PCPD\cite{sankaranarayanan2009path}, and CPD\cite{botea2013moving} adopt first move table. First move table has the similar feature like last move table, but it is suitable to solve point-to-point shortest path problem. When querying the shortest path from vertex s to vertex t, first move table can iteratively give the next vertex of the shortest path beginning from s. Our method is designed for SSSP problem, so the result is a tree, not a single path. A vertex in a tree may have several successors, thus the first move table, which is an array and can just store one successor of a vertex on a shortest path, cannot fit our requirement. If we persist in using first move table to record the tree, it need to be implement in the form of a data structure about tree, which is more expensive than an array.
We use last move table. A vertex in a tree only have one predecessor, so last move table can be stored in the form of an array. That is also why many classical algorithms for SSSP problem, like Dijkstra, use last move table to record the shortest path tree. So SPLZ adopts last move table as the form of storage of result.

Generally, SPLZ contains 3 parts: (1)calculating the APSP, (2)compressing the APSP offline, (3)decompressing the APSP online. We adopt PHAST algorithm to calculating the APSP. Other existing methods, like Dijkstra, Bellman-Ford-Moore, are also feasible. Next we focus on the compressing and decompressing method of SPLZ, which is a variant of LZ77.

\subsection{LZ77 algorithm}
Let $data$ be a string to be compressed, and $i$ bytes has been compressed. $DICT\_SIZE$ is the size of dictionary. The compression procedure of LZ77 is as Algorithm 1. LZ77 keeps the dictionary, $Dict$, sliding with $i$ increases.
$Longest\_match$ looks for the longest common subsequence among $dict+uncompressed data$ (begins from $dict$ and may extend into uncompressed data) and the prefix of uncompressed data, and then returns location and length of the common subsequence. Finally the compressed data is an array of $(location, length)$ pairs.

\begin{algorithm}
    \caption{The compression procedure of LZ77}
    \begin{algorithmic}[1]
        \While{(i$\leq$ len(data))}
        \State dict=data[i-DICT\_SIZE..i];
        \State (location, length)=longest\_match(dict+data[i..end], prefix of data[i..end]);
        \State output (location, length);
        \State i=i+length;
        \EndWhile
    \end{algorithmic}
\end{algorithm}

Let $compressed\_data$ be an array of $(location, length)$ pairs to be decompressed, and assume $i$ bytes has been decompressed. The decompression procedure of LZ77 is as Algorithm 2. Decompression is much simpler than compression. It successively loads every $(location, length)$ in compressed data, and then looks up and outputs the corresponding subsequence in $dict$. $dict$ still needs to slide with $i$.

\begin{algorithm}
    \caption{The decompression procedure of LZ77}
    \begin{algorithmic}[1]
        \ForAll{(location, length) in compressed\_data}
        \State dict=decompressed\_data[i-DICT\_SIZE..i];
        \State output dict[location..location+length];
        \State i=i+length;
        \EndFor
    \end{algorithmic}
\end{algorithm}

LZ77 is a representation of regular compression methods without any domain knowledge. Decompression speed of LZ family algorithm is fast, but decompressing some particular data from compressed data is dependent on previous data because of the sliding of dictionary. Other common LZ-family methods, like what are used in gzip, zlib and 7z, do not have any superiority on retrieval speed compared with LZ77, because they need to not only look up the dictionary to convert $(location, length)$ to original data and slice the dictionary, but also calculate some complex coding. Assuming that there are compressed data contains $n$ SPTs, from which we hope to decompress a particular SPT, we need decompress $\frac{n}{2}$ SPTs on average. This process results in a large amount of redundant operations.
\subsection{Fixed-dictionary compression}
To avoid redundant operations, we fix the dictionary. This means the dictionary is a fixed-size and fixed-location sequence in the front of the raw data. Data in the dictionary will not be compressed, to achieve a faster decompressing speed. These modification result in a lower compression ratio. We should point out that, SPLZ is not a compression algorithm for general situation, but an algorithm for solving SSSP problem.

Let $data$ be a set of SPT, and $data[0]$, the first SPT in $data$, be the dictionary. $Index[i]$ is the starting position of $i$-th SPT in compressed data. In other words, $index[i+1]-index[i]$ is equal to the compressed length of $i$-th SPT. Let $len(compressed\_data)$ be the total size of compressed data. Algorithm 3 describes the process of compressing. Firstly SPLZ loads the first SPT in $data$ as the dictionary and outputs it without compression. Then SPLZ compresses remaining SPTs successively. For each SPT, SPLZ repeatedly finds the longest common subsequence among $dict$ and the prefix of this SPT, and compresses, outputs and deletes the matched prefix, until this SPT is empty. There are many methods for finding the longest match \cite{bell1993longest}, we use a simple implement with a modification of KMP\cite{knuth1977fast}. Algorithm 4 is the process of decompressing the $m$-th SPT. $dict$ is not compressed so it can be loaded straightly. Next SPLZ locates the compressed $m$-th SPT in the compressed data stream using $index[m]$ and $index[m+1]$. For each $(location, length)$ pair, we can look up the corresponding subsequence in $dict$ to output the original data. The first line of Algorithm 3 and Algorithm 4 is just an assignment to a pointer, without copying any real data. In decompression process, the whole size of outputed data in line 4 is $\|V\|$.

\begin{algorithm}
    \caption{The compression procedure of SPLZ}
    \begin{algorithmic}[1]
        \State dict=data[0];
        \State output dict;
        \ForAll{spt in data-data[0]}
            \State index[number of spt in data]=len(compressed\_data);
            \While{(spt!=NULL)}
                \State (location£¬length)=longest\_match(subsequence of dict, prefix of spt);
                \State output (location, length);
                \State spt=spt-matched prefix of spt;
            \EndWhile
        \EndFor
    \end{algorithmic}
\end{algorithm}

\begin{algorithm}
    \caption{The decompression procedure of SPLZ}
    \begin{algorithmic}[1]
        \State dict=compressed\_data[0..sizeof(SPT)];
        \For {i=index[m] to index[m+1]}
            \State (location, length)=compressed\_data[i];	
            \State output dict[location..location+length];
        \EndFor
    \end{algorithmic}
\end{algorithm}

\subsection{An example of SPLZ compression and decompression}
Assume that there is a graph G as Figure 1 shows. After calculating shortest path tree for each vertex, we get six SPTs shown in Table \ref{tab:SPTs-raw}. If $SPT(V_i)[j]=k$, it means the precursory vertex of $V_j$ on the shortest path from $V_i$ to $V_j$ is $V_k$.

\begin{figure}
  \centering
  \includegraphics[width=.4\textwidth]{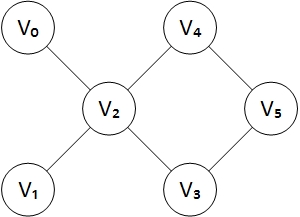}\\
  \caption{Graph G}\label{fig:example}
\end{figure}
\begin{table}
  \centering
  \begin{tabular}{ccccc}
     \noalign{\smallskip}\hline\noalign{\smallskip}
     \multirow{1}{*}{Vertex} &  \multicolumn{4}{c}{Adjacent vertices of $V_i$}\\
     \noalign{\smallskip}\hline\noalign{\smallskip}
     $V_0$ & $V_2$ &       &   &   \\
     $V_1$ & $V_2$ &       &   &   \\
     $V_2$ & $V_0$ & $V_1$ & $V_4$ & $V_5$ \\
     $V_3$ & $V_2$ & $V_5$ &   &   \\
     $V_4$ & $V_2$ & $V_5$ &   &   \\
     $V_5$ & $V_3$ & $V_4$ &   &   \\
     \hline\noalign{\smallskip}
   \end{tabular}
  \caption{The adjacency list of G}
  \label{tab:adjacency list}
\end{table}

\begin{table}
  \centering
  \begin{tabular}{ccccccc}
  \noalign{\smallskip}\hline\noalign{\smallskip}
  SPT  & \multicolumn{6}{c}{Elements of $SPT(V_i)$}\\
  \noalign{\smallskip}\hline\noalign{\smallskip}
  $SPT(V_0)$ & - & 2 & 0 & 2 & 2 & 3\\
  $SPT(V_1)$ & 2 & - & 1 & 2 & 2 & 3\\
  $SPT(V_2)$ & 2 & 2 & - & 2 & 2 & 3\\
  $SPT(V_3)$ & 2 & 2 & 3 & - & 2 & 3\\
  $SPT(V_4)$ & 2 & 2 & 4 & 2 & - & 4\\
  $SPT(V_5)$ & 2 & 2 & 4 & 5 & 5 & -\\
  \hline\noalign{\smallskip}
  \end{tabular}
  \caption{SPT of each vertex}
  \label{tab:SPTs-raw}
\end{table}
To obtain an effective compression, we convert Table 2 to another form. In Table 3, $SPT(V_i)[j]=k$ means the precursory vertex of $V_j$ on the shortest path from $V_i$ to $V_j$ is the $k$-th adjacent vertex of $V_j$. We define $SPT(V_i)[i]=0$.

Then, for example, we select $SPT(V_2)$ as the dictionary. Every SPT is compressed into an array of 2-tuple $(location, length)$. Note that sometimes a number in a SPT might not exist in the dictionary. For example, in Table \ref{tab:SPTs-converted}, $SPT(V_1)$[2]=1, but ``1'' does not exist in the dictionary. At this situation, we set $length$=0 and $location$=the number excluded in dictionary.

\begin{table}
  \centering
  \begin{tabular}{cccccccl}
  \noalign{\smallskip}\hline\noalign{\smallskip}
  SPT  & \multicolumn{6}{c}{Elements of $SPT(V_i)$} & After compressing\\
  \noalign{\smallskip}\hline\noalign{\smallskip}
  $SPT(V_0)$ & 0 & 0 & 0 & 0 & 0 & 0 & (0,6)\\
  $SPT(V_1)$ & 0 & 0 & 1 & 0 & 0 & 0 & (0,2) (1,0) (0,3)\\
  $SPT(V_2)$ & 0 & 0 & 0 & 0 & 0 & 0 & dictionary\\
  $SPT(V_3)$ & 0 & 0 & 2 & 0 & 0 & 0 & (0,2) (2,0) (0,3)\\
  $SPT(V_4)$ & 0 & 0 & 3 & 0 & 0 & 1 & (0,2) (3,0) (0,2) (1,0)\\
  $SPT(V_5)$ & 0 & 0 & 2 & 1 & 1 & 0 & (0,2) (2,0) (1,0) (1,0) (0,1)\\
  \hline\noalign{\smallskip}
  \end{tabular}
  \caption{ SPT of each vertex after converting and compressing}
  \label{tab:SPTs-converted}
\end{table}

To make a simple illustration, here we assume that each 2-tuple needs two bytes. Therefore after compressing, the index array is:
(0, 2, 8, 14, 20, 28, 38).

The total length of output is 44 bytes, for the length of dictionary is 6 bytes and the length of compressed date is 38 bytes. The effectiveness of compression seems poor, because the scale of graph in our example is too small.

When solving a SSSP problem online, for example, calculating the $SPT(V_1)$, the steps are:
\begin{enumerate}
  \item Find out that $index[1]=2$ and $index[2]-index[1]=6$. In other words, the length of compressed $SPT(V_1)$ is 6 and its start location in whole compressed data is 2.
  \item Convert every 2-tuple to original data by looking up the dictionary. For example, when handling the 2-tuple (0, 2), we intercept the subsequence of dictionary, which begins at 0 and is of length 2. \\
      All the conversion is: (0, 2) $\rightarrow$ 0 0; (1, 0) $\rightarrow$ 1; (0, 3) $\rightarrow$ 0 0 0. \\
      This step can be easily parallelized.
  \item Concatenate these subsequences to one array: (0 0 1 0 0 0). This array is $SPT(V_1)$.
\end{enumerate}

\section{Details of implementation}
\label{sec:Improvements}
\subsection{The key factor affecting the compression ratio}
When we use fixed-dictionary compression, the compression ratio is mainly decided by the similarity between the dictionary and data to be compressed. Let $path\textnormal{-}len(u, v)$ be the number of edges along the shortest path between vertex $u$ and vertex $v$. Similarity between two SPTs has a negative correlation with $path\textnormal{-}len(u, v)$ between the source vertex of the two SPTs. We use the proportion of common edges among two SPTs to measure the similarity between two SPTs.

\begin{figure}
\centering
\includegraphics[width=.6\textwidth]{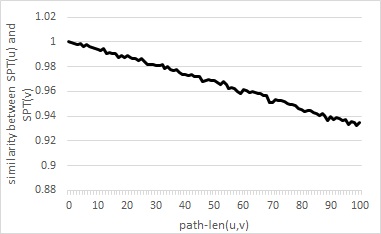}
\caption{Relation between similarity and path-len}
\label{fig:sim-path-len}
\end{figure}

Figure 2 presents an experiment result, which is based on a northwest USA road network with about 1.2 million vertices. . The less $path\textnormal{-}len(u, v)$ of two vertex $u$ and $v$, the higer similarity between $SPT(u)$ and $SPT(v)$. This result is reasonable in real-world. For example, assume there are three locations A, B, and C. Both the distance of (A, C) and (B, C) is 10 km. If the distance between A and B is one meter, we could guess that the shortest path (A, C) and (B, C) are almost the same. When we choose $SPT(u)$ as dictionary and compress $SPT(v)$, the impact of $path\textnormal{-}len(u, v)$ on the compression ratio is as Figure 3 shows.

In addition, Figure 2 also shows how frequently path coherence holds. When $path\textnormal{-}len(u, v)$ is less than 100 (vertices contained in a coherent region), $SPT(u)$ and $SPT(v)$ share over 93\% of their elements. An element in $SPT(u)$ records the last move of the path from $u$ to a vertex. This experiment validates that path coherence holds in most cases on a large-scale road network.

\begin{figure}
\centering
\includegraphics[width=.6\textwidth]{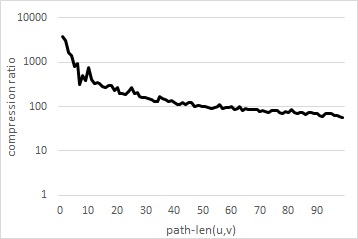}
\caption{Relation between compression ratio and path-len}
\label{fig:ratio-path-len}
\end{figure}
The result points out that the compression ratio decreases fast with increasing $path\textnormal{-}len(u, v)$. To reduce the space consumption, it is necessary to limit the $path\textnormal{-}len(u, v)$ between the dictionary and the SPT to be compressed.
\subsection{Regions partition}
If we choose only one SPT as the dictionary in a large scale graph, there always are many vertices far away from the dictionary. By partitioning the graph into a series of regions with smaller size, we can choose a SPT as dictionary and compress the rest SPT independently for every region. We choose the SPT of a vertex which is closest to the geometrical center as the dictionary, and call this vertex the root of this region. Partition ensures that in a region, the path-len between vertex of dictionary and other vertices don't exceed the diameter of the region.

When partitioning the graph into numerous disjoint regions, we should keep the distance of vertices among a region as close as possible. It can be handled as a clustering problem. We use k-means, a simple but effective clustering method, to partition the graph. The simplest attribute for clustering vertices in a road network is coordinate, and we adopt it. Actually, there are numerous methods to partition the graph without coordinate. Coordinate is an unessential condition for SPLZ.

Due to that data in the dictionary will not be compressed and output in raw form, if the number of regions is excessive, uncompressed data would occupy a large proportion in final output. If the number of regions is small, we cannot ensure that the diameter of a region is significantly less than the diameter of the total graph. Intuitively, to reach both small number of regions and small size of every region, we assume that optimal number of regions is in form of $C*\sqrt{\|V\|}$, and choose a proper value of $C$ by experiment. Section 5.2 compares the impact of different parameter $C$.

\subsection{Multi-step compression}
\label{subsec:multi-step}
Assume that $u$ is a root of a region, and vertices of SPT($u$) among its region is as Figure 4 shows.
\begin{figure}
\centering
\includegraphics[width=.5\textwidth]{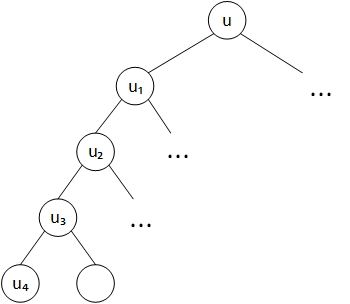}
\caption{Part of SPT(u) within a region}
\label{fig:multi-step-example}
\end{figure}
In region showed in Figure 4, all SPT except $SPT(u)$ is compressed with dictionary $SPT(u)$. We call this process one-step compression. Path-len between most SPT in this region and $SPT(u)$ is 3 or 4.

We can reduce the path-len by multi-step compression. For example, let the grandparent of each SPT be its dictionary. As for vertex $u_4$, $SPT(u_4)$ will be compressed with dictionary $SPT(u_2)$ and $SPT(u_2)$ will be compressed with dictionary $SPT(u)$. When we decompress $SPT(u_4)$, we must decompress $SPT(u_2)$ at first. It is so-called two-step compression for vertex $u_4$. By applying similar operation to all vertex, their path-len is decreased to 1 or 2.

We call the path-len between a SPT and its dictionary $len\textnormal{-}to\textnormal{-}dic$. Figure 3 tell us that shorter path-len leads to a higher compression ratio. But the reduction of space costs brings higher time cost. If $len\textnormal{-}to\textnormal{-}dic$ is $d$, the time to decompress $SPT(u_4)$ is $\lceil path\textnormal{-}len(u,u_4)/d\rceil$ times of one-step compression. By controlling $len\textnormal{-}to\textnormal{-}dic$, we can adjust the point of balance between space costs of compression and time costs of decompression online.

We define $len\textnormal{-}to\textnormal{-}dic=\infty$ when we use one-step compression.
\subsection{Code of compressed data}
The compressed data are array of $(location, length)$. If $location$ and $length$ are fixed-length integer, data compressed by method in Section 3.3 can be compressed one more time by entropy coding. However, entropy coding has poor decompression speed. We adopt a variable length coding to encode $location$ and $length$. Though our coding method cannot obtain the compression ratio as high as entropy coding, it has almost no negative effect on decompression speed. The code is also prefix coding, so there is no ambiguity when we decode it.

In the data stream of compressed data, $location$ is presented by differential coding. We just record the difference between every $location$ and its predecessor except the first one, because difference between two adjacent $location$ usually smaller than their real value. Differential coding might result in a shorter code. Value of $length$ does not have such a feature, so we record its real value.

\begin{table}
\begin{tabular}{lll}
\noalign{\smallskip}\hline\noalign{\smallskip}
range of value (in hex) & code length & code format(in binary)  \\
\noalign{\smallskip}\hline\noalign{\smallskip}
[0x00, 0x0F] (uncompressed) & 1 & 1111xxxx \\\relax
[0x00, 0x7F] & 1 & 0xxxxxxx\\\relax
[0x0080, 0x3FFF]& 2 & 10xxxxxx xxxxxxxx  \\\relax
[0x004000, 0x1FFFFF]& 3 & 110xxxxx xxxxxxxx xxxxxxxx \\\relax
[0x00200000, 0x0FFFFFFF]& 4 & 1110xxxx xxxxxxxx xxxxxxxx xxxxxxxx \\
\noalign{\smallskip}\hline
\end{tabular}
\caption{Code for length}
\label{tab:code of length}
\end{table}

\begin{table}
\begin{tabular}{lll}
\noalign{\smallskip}\hline\noalign{\smallskip}
range of value (in hex) & code length & code format(in binary)  \\
\noalign{\smallskip}\hline\noalign{\smallskip}
[0x00, 0x3F] & 1 & 00xxxxxx \\\relax
[-0x00, -0x3F] & 1 & 01xxxxxx\\\relax
[0x0040, 0x1FFF] & 2 & 100xxxxx xxxxxxxx  \\\relax
[-0x0040, -0x1FFF] & 2 & 101xxxxx xxxxxxxx \\\relax
[0x002000, 0x0FFFFF] & 3 & 1100xxxx xxxxxxxx xxxxxxxx \\\relax
[-0x002000, -0x0FFFFF] & 3 & 1101xxxx xxxxxxxx xxxxxxxx \\\relax
[0x00100000, 0x07FFFFFF] & 4 & 11100xxx xxxxxxxx xxxxxxxx xxxxxxxx \\\relax
[-0x00100000, -0x07FFFFFF] & 4 & 11101xxx xxxxxxxx xxxxxxxx xxxxxxxx \\
\noalign{\smallskip}\hline
\end{tabular}
\caption{Code for location}
\label{tab:code of location}
\end{table}
Encoding method for $length$ and $location$ in detail is separately in Table 4 and Table 5. In the first line of Table 4, ``uncompressed'' means that some bytes does not appear in the dictionary, so these bytes cannot be compressed. In our method, the value of such a byte must be no more than 15. It is reasonable for a real-world road network, for number of branch of real-world road usually smaller than 15. In the case of that degree of a vertex $u$ is more than 15, we can add a virtual vertex $u\prime$ to the graph. Let the distance between $u$ and $u\prime$ be zero and assign excess edges of $u$ to $u\prime$.

\section{Experiment}
\label{sec:Experiment}
\subsection{Experiment setup}
Our experiment code is written in C++, and compiled by VC++ 2010. The program includes two parts: precomputing offline and calculating the SSSP online. The experiments run on a PC, with 3.4 GHz Intel i7-4770(4 cores), 24GB RAM. External memories include a 2TB mechanical disk and a 256GB SSD. For parallelly precomputing, we use OpenMP. Data of graph are downloaded from http://www.dis.uniroma1.it/$\sim$challenge9, which are benchmarks for the 9th DIMACS Implementation Challenge\cite{demetrescu2009shortest}. The data set we used is "Northwest USA"(NW), with 1207945 vertices and 2840208 edges, and the type of graph is "Distance graph".

The source code of our experiments is released\footnote{https://github.com/asds25810/SPLZ}.

\subsection{Precomputing}
Operation of precomputing consist of computing the APSP and compressing it. The target of compressing is to reduce the space consumption of APSP. So the compression ratio is a important feature for measuring the effectiveness of precomputing. The number of regions has an impact on compression ratio. We choose different setting of parameter $C$ for $C*\sqrt{\|V\|}$ as the number of regions separately. For every parameter setting, the running time of precomputing is about 13 hours. Table 6 shows the effect of number of regions on the compression ratio.

The number of regions determine the average size of each region, and the size of a region has effect on the path-len between vertices in a region. The less the path-len between vertices, the higher the compression ratio. So it seems that more number of regions may lead to higher compression ratio. However, Table 6 demonstrates that when the number of regions is more than a certain value, the compression ratio falls down. It is owing to that, with the number of regions increases, the proportion of dictionary increases. We select a vertex as the representative vertex for each region. The SPT of the representative vertex is the dictionary of that region. To ensure that the dictionary is available at the immediate time of decompressing, dictionary will not be compressed. Although more number of regions leads to a higher compression ratio of single SPT, the total size of final data will increase because the increased size of dictionaries.

If the dictionaries occupy a high proportion of the final output, we can consider to compress the dictionaries. But it will result in two problems. One is the extra time cost for decompressing dictionary when decompressing data. Another one is that, actually, it is difficult to find a proper "dictionary" for compressing dictionaries, which intrinsically have less data-redundancy. In other words, the compression ratio of compressing the dictionaries is much lower than compressing SPTs.

We set $1*\sqrt{\|V\|}$ as the number of regions for following experiment.

\begin{table}
  \centering
  \begin{tabular}{cccccc}\\
     \noalign{\smallskip}\hline\noalign{\smallskip}
     \tabincell{c}{C} & \tabincell{c}{Raw size \\ $\left[GB\right]$}& \tabincell{c}{Compressed \\ size$\left[GB\right]$} & \tabincell{c}{Compression \\ ratio} & \tabincell{c}{Dictionary size \\$\left[GB\right]$ }& \tabincell{c}{Proportion of \\ dictionary} \\
     \noalign{\smallskip}\hline\noalign{\smallskip}
      0.5&1459&16.65& 88& 0.66& 4.0\%\\
        1&1459&13.29&110& 1.33&10.0\%\\
        2&1459&11.75&124& 2.65&22.5\%\\
        4&1459&12.21&119& 5.31&43.5\%\\
        8&1459&15.87& 92&10.62&66.9\%\\
     \noalign{\smallskip}\hline
   \end{tabular}
  \caption{Effect of number of regions on compression ratio}
  \label{tab:number of regions}
\end{table}

\subsection{Multi-step compression}
\begin{figure}
  \centering
  \includegraphics[width=.7\textwidth]{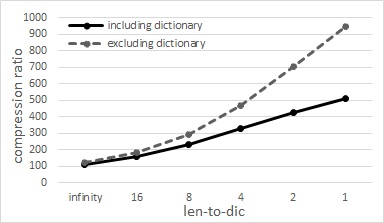}\\
  \caption{Effect of Len-to-dic on compression ratio}
  \label{fig:len-to-dic}
\end{figure}

Table \ref{tab:number of regions} shows the results of one-step compressing with different number of regions. Multi-step compressing can achieve a higher compression ratio as shown in Figure \ref{fig:len-to-dic}. SPLZ can adjust the compression ratio by controlling the parameter $len\textnormal{-}to\textnormal{-}dic$, which was described in Section \ref{subsec:multi-step}. This capability of SPLZ makes it adaptable to different capacity of memory. With the $len\textnormal{-}to\textnormal{-}dic$ decreases, the compression ratio increases. It is due to the similarity between the SPT to be compressed and its dictionary is higher when the parameter $len\textnormal{-}to\textnormal{-}dic$ decreasing. SPLZ achieves the highest compression ratio when $len\textnormal{-}to\textnormal{-}dic$=1. The APSP of size about 1459 GB can be compressed to 2.87 GB. It is affordable for an ordinary PC.

\subsection{Online performance}
\label{subsec:performance}
After precomputing, we test the time costs of solving the SSSP problem for a particular vertex. We load the compressed APSP in memory, and randomly generate a series of queries. Each query input a vertex $v$, and request $SPT(v)$ as output. Table \ref{tab:time costs} shows the average costs of handling a query. The parameter $len\textnormal{-}to\textnormal{-}dic$ evidently affects the performance. Higher $len\textnormal{-}to\textnormal{-}dic$ makes a lower space consumption. However, the reduced space consumption is repayed by increasing time costs. $Len\textnormal{-}to\textnormal{-}dic$ should be decided by the bottleneck of different applications.

If the memory capacity is not enough to store the compressed APSP, we can consider to store it to external memory. When SPLZ handles a query, it looks up the compressed SPT from external memory, and then decompresses it and returns the result. Compared with RAM, external memory is cheaper and usually has higher capacity. The time costs on external memory in Table \ref{tab:time costs} is the stable performance after handling a large number of random queries. The latency of accessing mechanical disk is significantly higher than that of memory. When $len\textnormal{-}to\textnormal{-}dic$ is not $\infty$, the results on mechanical disk have obvious difference in  each experiment execution, and we do not find any regular pattern. So we only record the result when $len\textnormal{-}to\textnormal{-}dic$ is $\infty$.  SSD performs better than mechanical disk. The time cost on SSD is three to four times as much as that on memory. Note that the space costs only consider the compressed APSP. Other consumptions are much less than them.

If the parent pointers need to be converted to the global id of each vertices in results, another 887${\mu}s$ is needed. This conversion is not always needed in our opinion. To represent a path tree in parent pointer form, the local id (represented like Table 3) may be enough if graph is stored in the form of adjacency list.

\begin{table}
  \centering
  \begin{tabular}{ccccc}
     \noalign{\smallskip}\hline\noalign{\smallskip}
     Methods & Len-to-dic & Time (${\mu}s$) &  \tabincell{c}{Space(RAM)\\ $\left[GB\right]$ }& \tabincell{c}{Space(external)\\ $\left[GB\right]$}\\
     \noalign{\smallskip}\hline\noalign{\smallskip}
     \multirow{6}{*}{SPLZ on memory} & $\infty$ & 219 &  13.29 & -\\
       &16 & 323 & 9.37 & -\\
       & 8 & 464 & 6.32 & - \\
       & 4 & 733 & 4.44 & - \\
       & 2 & 1265 & 3.41 & -  \\
       & 1 & 2287 & 2.87 & -  \\
     \noalign{\smallskip}\hline\noalign{\smallskip}

     \multirow{6}{*}{SPLZ on SSD}   & $\infty$ &  740  & 1.33 & 11.96  \\
       &16 &  1176 & 1.33 & 8.04 \\
       & 8 &  1709 & 1.33 & 4.99 \\
       & 4 &  2556 & 1.33 & 3.11 \\
       & 2 &  4468 & 1.33 & 2.08  \\
       & 1 &  8002 & 1.33 & 1.54  \\
    \noalign{\smallskip}\hline\noalign{\smallskip}
    SPLZ on mechanical disk & $\infty$ &  7085  & 1.33 & 11.96  \\
    \noalign{\smallskip}\hline\noalign{\smallskip}
    Dijkstra & - & 217291 & - & -\\
    \noalign{\smallskip}\hline
   \end{tabular}
  \caption{Time and space cost online}
  \label{tab:time costs}
\end{table}

Although SPLZ needs about 13 GB space costs if $len\textnormal{-}to\textnormal{-}dic$=$\infty$, storing the compressed APSP into disk or selecting a lower $len\textnormal{-}to\textnormal{-}dic$ would make the space consumption more practicable at the expense of lower quering.

The average time costs of Dijkstra algorithm based on binary heap is about 217 ms on our experiment graph. In our experiment, the performance of SPLZ is almost three orders of magnitude faster than Dijkstra based on binary heap, if let $len\textnormal{-}to\textnormal{-}dic$=$\infty$. We also implement PHAST, which needs 27.4ms in our experiment. These comparisons may be influented by the details of how a people implements these algorithm. To fairly shows the online performance of SPLZ, we try to find a lower bound of running time of SSSP problem, and compare SPLZ with this lower bound.

\subsection{Lower bound of SSSP problem}
The time costs of solving SSSP problem has a natural lower bound. Whatever methods we use to calculate the shortest path from a vertex to all vertices in a graph, we must fill the result to an array of length $\|V\|$ as output. So the natural lower bound is the time costs of copying an array of length $\|V\|$. We compares the time costs of SPLZ and array copying in Table \ref{tab:lower bound}. We test two copying methods: memcpy and for-loop assignment. Memcpy() is a standard function in C library, which is fully optimized. Considering many algorithm successively output their result in a loop, we also test copying an array by for-loop(assigning the elements one by one). The results in Table \ref{tab:lower bound} shows that the performance of SPLZ is close to the lower bound.

\begin{table}
  \centering
  \begin{tabular}{ll}
    \noalign{\smallskip}\hline\noalign{\smallskip}
    Methods & Time costs(${\mu}s$) \\
    \noalign{\smallskip}\hline\noalign{\smallskip}
    SPLZ & 219 \\
    memcpy & 84 \\
    for-loop assignment & 448 \\
    \noalign{\smallskip}\hline
  \end{tabular}
  \caption{Time costs of solving a SSSP problem by SPLZ and copying an array of length $\|V\|$}
  \label{tab:lower bound}
\end{table}

\subsection{Experiments on other road networks}
Up to now, we test SPLZ on only one graph, the Northwest USA road network. Here we show the results of experiments on some other graph data. Table \ref{tab:more data} introduces the related information about these data. Like North west USA data, these data are also downloaded from http://www.dis.uniroma1.it/$\sim$challenge9.
These experiments are used to show the performance of SPLZ on different data. Here we set $C=1$ and $len\textnormal{-}to\textnormal{-}dic=\infty$. Other hardware and software configurations are the same to section 5.1.

Table \ref{tab:more data offline} shows the results of preprocess of SPLZ on these datasets. The scale of FLA dataset is close to Northwest, so the performance of SPLZ is similar to what we have shown above. SPLZ achieves less compression ratio on smaller graph, because path coherence takes effects when vertices in a coherent region are sufficiently far away from a vertex, while the average distance between vertices in smaller graph is closer. Table \ref{tab:more data online} is the decompression performance on these datasets. The lower bound cost is also tested. Table \ref{tab:more data online} validates that SPLZ also works well on small-scale road networks.

\begin{table}
  \centering
  \begin{tabular}{llcc}
     \noalign{\smallskip}\hline\noalign{\smallskip}
     Name & Description & Number of vertices & Number of edges \\
     \noalign{\smallskip}\hline\noalign{\smallskip}
       FLA &  Florida & 1,070,376 & 2,712,798 \\
       COL &  Colorado & 435,666 & 1,057,066 \\
       BAY &  San Francisco Bay Area & 321,270 & 800,172  \\
    \noalign{\smallskip}\hline
   \end{tabular}
  \caption{Information about more datasets.}
  \label{tab:more data}
\end{table}

\begin{table}
  \centering
  \begin{tabular}{ccccc}
     \noalign{\smallskip}\hline\noalign{\smallskip}
     Name &  \tabincell{c}{Raw size\\$\left[GB\right]$ } &  \tabincell{c}{Compressed size\\$\left[GB\right]$ } & Compression ratio&\tabincell{c}{Preprocess time\\$\left[h:m\right]$} \\
     \noalign{\smallskip}\hline\noalign{\smallskip}
       FLA &  1146 & 9.97 & 115 & 10:56 \\
       COL &  190  & 2.72 & 70  & 2:15 \\
       BAY &  103  & 1.42 & 73  & 1:07  \\
    \noalign{\smallskip}\hline
   \end{tabular}
  \caption{The compression ratio and time cost on more datasets.}
  \label{tab:more data offline}
\end{table}

\begin{table}
  \centering
  \begin{tabular}{clccc}
     \noalign{\smallskip}\hline\noalign{\smallskip}
     Graph & Methods &  Time (${\mu}s$) &  \tabincell{c}{Space(RAM)\\$\left[GB\right]$} &  \tabincell{c}{Space(external)\\$\left[GB\right]$}\\
     \noalign{\smallskip}\hline\noalign{\smallskip}
     \multirow{5}{*}{FLA} & SPLZ on memory & 172 &  9.97 & -\\
                          & SPLZ on mechanical disk & 6920 & 1.11& 8.86\\
                          & SPLZ on SSD& 657& 1.11& 8.86\\
                          & memcpy& 76 & - & - \\
                          & for-loop assignment& 459 & - & - \\
     \noalign{\smallskip}\hline\noalign{\smallskip}
     \multirow{5}{*}{COL} & SPLZ on memory & 98 &  2.72 & -\\
                          & SPLZ on mechanical disk & 6120 &0.29 &2.43\\
                          & SPLZ on SSD & 449 & 0.29&2.43\\
                          & memcpy & 29 & - & -\\
                          & for-loop assignment & 192 & - & -\\
    \noalign{\smallskip}\hline\noalign{\smallskip}
    \multirow{5}{*}{BAY} & SPLZ on memory & 71 &  1.42 & -\\
                          & SPLZ on mechanical disk & 5767 &0.18 &1.24\\
                          & SPLZ on SSD & 371 & 0.18&1.24\\
                          & memcpy & 21 & - & -\\
                          & for-loop assignment & 140 & - & - \\
    \noalign{\smallskip}\hline
   \end{tabular}
  \caption{Time and space cost online on more datasets}
  \label{tab:more data online}
\end{table}

\section{Conclusion}

\label{sec:Conclusion}
In this paper, we presented SPLZ, an algorithm for solving single source shortest path problem on road network. SPLZ is about three orders of magnitude faster than Dijkstra based on binary heap. Compared with the time costs of array copying, which is a natural lower bound of SSSP problem, SPLZ shows a significantly high performance online. Even though SPLZ consumes much memory yet, this problem can be solved by storing compressed data into external memory or by adjusing the parameter $len\textnormal{-}to\textnormal{-}dic$.

Future research will focus on developing a more efficient preprocess method. In our experiments, SPLZ can solve SSSP prolem on a road network with about 1.2 millions vertices, and it is enough for many applications. But we should admit that, SPLZ still cannot deal with a more large-scale road network, because of the huge time costs for precomputing. We can make efforts to two points. One is to adopt an algorithm faster than sequential PHAST to calculate APSP, and the other is to use a more efficient methods to find the longest match while compressing the APSP.

\section*{Acknowledgement}
We would like to thank the reviewers for their valuable suggestions, and Shiyan Zhan for the fruitful discussions.
This work is supported by Natural Science Foundation of China (No. 61033009 and No. 61303047) and Anhui Provincial Natural Science Foundation (No. 1208085QF106).

\bibliographystyle{spbasic}      
\bibliography{reference}   


\end{document}